\documentclass[12pt]{article}
\usepackage{amssymb}
\usepackage{graphicx}

\newcommand{\be}{\begin{equation}}
\newcommand{\ee}{\end{equation}}
\newcommand{\bea}{\begin{eqnarray}}
\newcommand{\eea}{\end{eqnarray}}

\begin{document}

\begin{center}
\textbf{Piezocaloric and multicaloric effect in the KH$_2$PO$_4$ type ferroelectrics}
\end{center}

\begin{center}
A.S.Vdovych$^a$, A.P.Moina$^a$, R.R.Levitskii$^a$, I.R.Zachek$^b$
\end{center}

\begin{center}
$^a$\textit{Institute for Condensed Matter Physics, National Academy of Sciences of Ukraine\\
79011, 1 Svientsitskii Street, Lviv, Ukraine\\
$^b$ Lviv Polytechnic National
University, 12 Bandera Str., 79013 Lviv, Ukraine }
\end{center}

\sloppy

Using the proton ordering model modified by taking into account
the dependence of the dipole moments on the order parameter
[Vdovych {\it et al}, 2014], we explore the piezocaloric and
multicaloric effects in the  KH$_{2}$PO$_{4}$ type ferroelectrics,
caused by the shear stress $\sigma_6$ and longitudinal electric
field $E_3$. The multicaloric effect is shown to be stronger than
either electrocaloric or piezocaloric effects, especially at
temperatures far from the phase transition.



\section{Introduction}
The piezocaloric effect is a change of the sample temperature
$\Delta T$ at adiabatic changing of the applied to it mechanical
stress. This effect, just like the electrocaloric effect, is
explored because of its potential application in
environment-friendly and compact solid-state cooling devices.

The largest piezocaloric effect so far has been observed in shape memory alloys, undergoing a martensitic transformation.
Thus, in NiTi it was obtained that $\Delta T=17$~K under the tensile stress of $\sigma\thickapprox650$~MPa \cite{Cui2012} and
$\Delta T=40$~K at $\sigma\thickapprox800$~MPa \cite{Pieczyska2006}. In Cu$_{68.13}$Zn$_{15.74}$Al$_{16.13}$ the value of $\Delta T=15$~K was
reached  \cite{Bonnot2008}.

Ferroelastics are promising materials too. The combined
electrocaloric and piezocaloric effects (a multicaloric effect)
have been studied
 \cite{Lisenkov2013} by the ab initio method in PbTiO$_{3}$ crystals, which are both ferroelectric and ferroelastic. The calculated temperature
 change in large fields and stresses exceeds 30~K.

The hydrogen bonded ferroelectrics of the KH$_{2}$PO$_{4}$ (KDP) family are also ferroelastic. In the
ferroelectric phase the spontaneous polarization $P_3$ is accompanied by the spontaneous shear strain $\varepsilon_6$.
So far, neither the piezocaloric nor multicaloric effcts in these crystal have been explored yet.

Influence of the shear stress $\sigma_6$ and electric field $E_3$ on polarization, dielectric permittivity, piezomodules, elastic
constants of the KH$_{2}$PO$_{4}$ family crystals was described in \cite{Stasyuk2000,Stasyuk2001,lis2007,JPS1701} using the proton ordering model
 modified by taking into account a piezoelectric coupling to the shear strain $\varepsilon_6$. Shifts of the transition temperature and eventual smearing
  out
of the first order transition by these fields conjugate to the
order parameter were also described. However, these theories
suffered from the inner logical contradiction, associated with the
required invoking of two different values of the effective dipole
moments for the paraelectric and ferroelectric phases
\cite{Stasyuk2001,JPS1701}. Hence, while  no physical
characteristic of the crystal should exhibit any discontinuity in
the fields above the critical one, there was no smooth transition
between the values of model parameters, set to be different for
the two phases. This contradiction has been removed in
\cite{Vdovych2014prep,Vdovych2014} by taking into account the
assumed dependence of the effective dipole moment on the order
parameter. The term
\be - \left(\sum\frac{\sigma_{q'f'}}{2}\right)^2 \mu'E_3
\sum\frac{\sigma_{qf}}{2} \label{H_E}\ee has been introduced to
the Hamiltonian, equivalent to a term proportional to $P_3^3E_3$
in a phenomenological thermodynamic potential. Such a modification
allowed us \cite{Vdovych2014prep,Vdovych2014} to quantitatively
and consistently describe the behavior of the physical
characteristics of the KH$_{2}$PO$_{4}$ family crystals in
presence of the electric field $E_3$, including the electrocaloric
effect.

In this paper we use the model \cite{Vdovych2014} to explore the
piezocaloric and multicaloric effects in the  KH$_{2}$PO$_{4}$
type crystals as a reaction to the shear stress $\sigma_6$.

\section{Calculations}
No modification to the model developed in \cite{Vdovych2014} is
required to study the influence of $\sigma_6$ stress. All the
derived expressions for the entropy, polarization, and shear
strain $\varepsilon_6$ remain valid and can be found in
\cite{Vdovych2014prep,Vdovych2014}.

\begin{figure}[!h]
\begin{center}
 \includegraphics[scale=0.7]{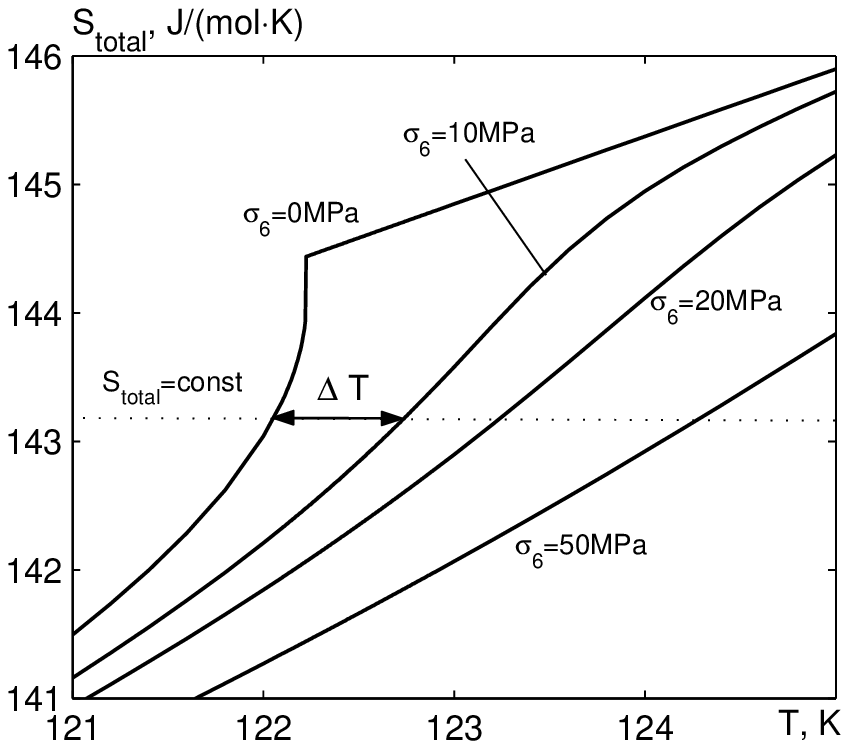} ~~~~ \includegraphics[scale=0.7]{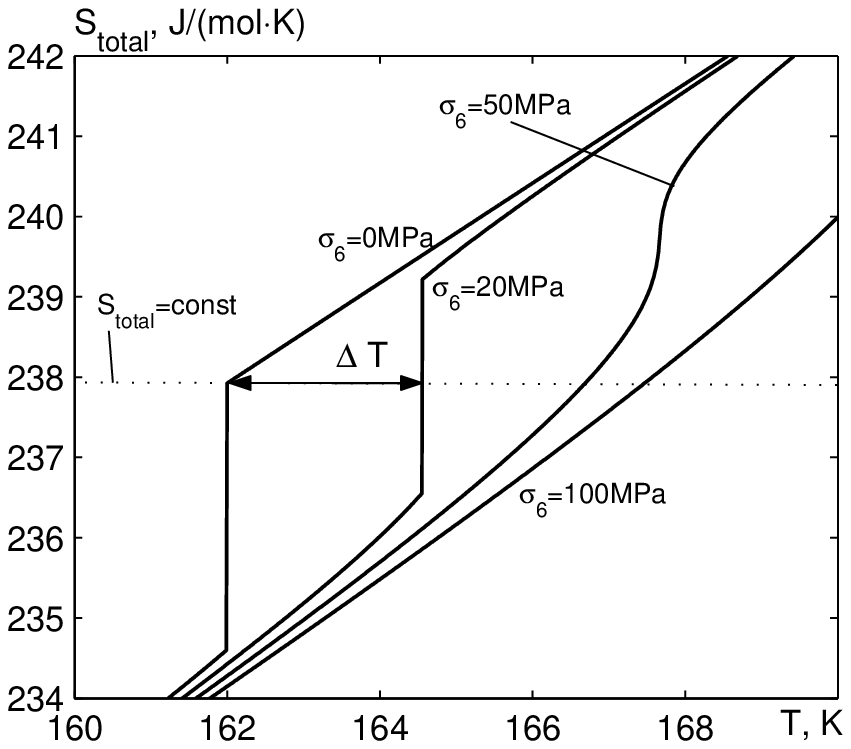}
\end{center}
\caption[]{The temperature dependence of molar entropy of KH$_{2}$PO$_{4}$ (left) and KD$_{2}$AsO$_{4}$ (right)
at different values of the stress $\sigma_6$.} \label{S_10MPa}
\end{figure}

From the entropy we obtain the molar specific of the proton subsystem $\Delta C^\sigma = T(\partial S/\partial T)_{\sigma}$.
%%  4.3
% \be \Delta
%C^\sigma = T\left( \frac{\partial S}{\partial T}\right)_{\sigma}, \ee
%%
The total specific heat is considered to be the sum  of  $\Delta
C^\sigma$ and of a some regular background contribution, mostly
from the lattice of heavy ions
\be C=\Delta C^\sigma + C_{reg}  \ee
The regular contribution near $T_c$ is approximated by the linear dependence
\[ C_{reg} = C_0+C_1(T-T_c), \]
yielding a good agreement with experimental data \cite{Vdovych2014prep,Vdovych2014}.
The corresponding regular contribution to the entropy  near  $T_c$ is then
\[ S_{reg} = \int \frac{C_{reg}}{T}dT = (C_0-C_1T_c)\ln(T)+C_1T +
const \]
Hence, the total entropy as a function of the field  and stress $\sigma_6$ is
\be S_{total}(T,E,\sigma_6)=S + S_{reg}. \label{Stotal} \ee
Solving (\ref{Stotal}) with respect to temperature at $S_{total}(T,E,\sigma_6)=const$ and two values of the
field and stress, one can calculate the electrocaloric, piezocaloric, or combined multicaloric temperature shifts
 (as seen in fig.~\ref{S_10MPa})
\be \Delta T = T(S_{total},E(2),\sigma_6(2))-T(S_{total},E(1),\sigma_6(1)). \label{DT_S} \ee

Numerical calculations are performed for the K(H$_{1-x}$D$_{x})_{2}$PO$_{4}$ crystals (at $x=0$ undergoing the first order phase
transition close to the second order transition)  and
K(H$_{1-x}$D$_{x})_{2}$AsO$_{4}$ (with a pronounced first order transition at $x=1$).
For this we use the same values of the model parameters that were obtained in \cite{Vdovych2014prep,Vdovych2014}.

\begin{figure}[!h]
\begin{center}
 \includegraphics[scale=0.625]{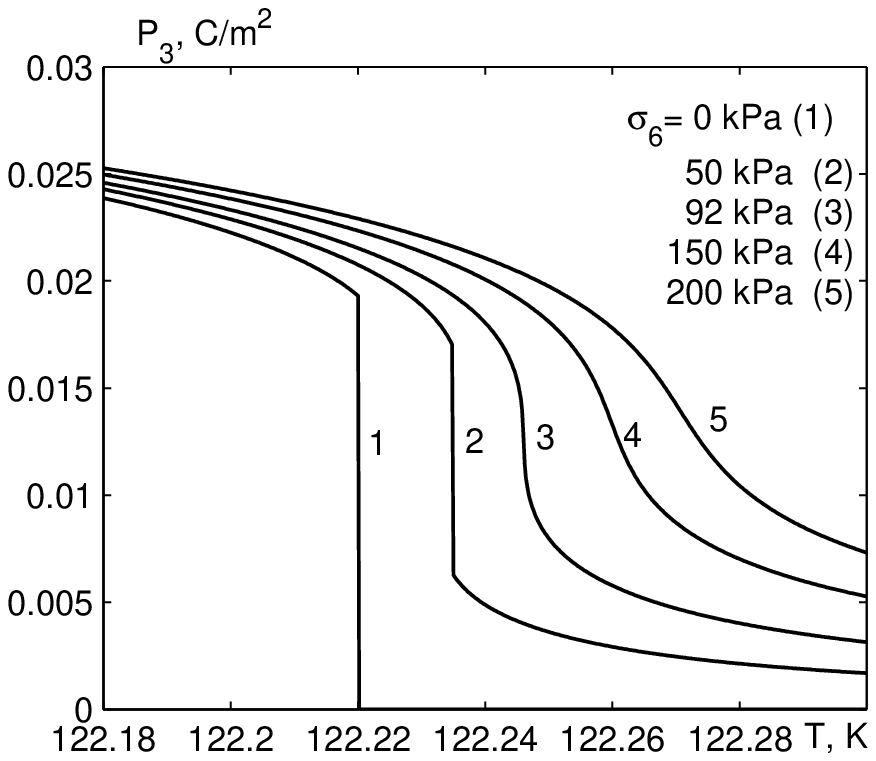}  \includegraphics[scale=0.625]{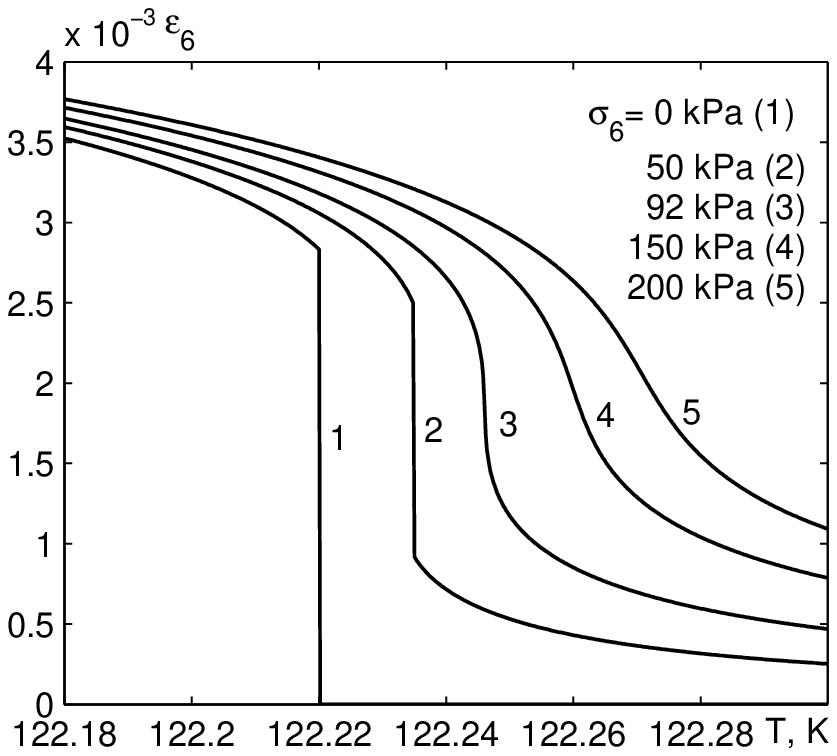}
\end{center}
\caption[]{The temperature dependence of polarization and shear
strain $\varepsilon_6$ of KH$_{2}$PO$_{4}$ at different values of
the stress $\sigma_6$.} \label{Ps_92kPa}
\end{figure}

Behavior of molar entropy (fig.~\ref{S_10MPa}), polarization  and
shear strain $\varepsilon_6$ (fig.~\ref{Ps_92kPa})  is typical for
the systems with the first order phase transitions in external
fields conjugate to the order parameter. The stress $\sigma_6$
induces non-zero strain $\varepsilon_{6}$ and polarization $P_{3}$
above the transition temperature. At low stresses both have jumps
at $T_c$, indicating the first order phase transitions. The jumps
decrease with increasing stress, and $T_c$ increases. Above the
certain critical stress $\sigma^*_6$ (the corresponding critical
temperature is $T^*_c$) the phase transition is smeared out. The
critical stress and temperature are  $\sigma^*_6=92$~kPa,
$T^*_c=122.246$~K for $x=0$ and $\sigma^*_6=6.2$~MPa,
$T^*_c=212.585$~K for $x=0.89$.

\begin{figure}[!h]
\begin{center}
 \includegraphics[scale=0.65]{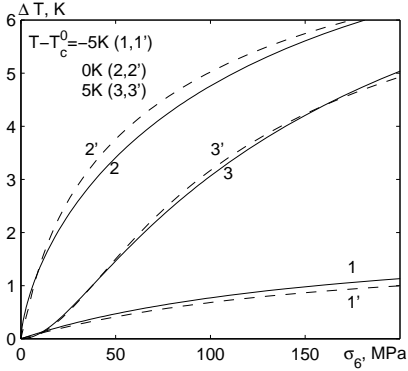}
\end{center}
\caption[]{The stress $\sigma_6$ dependence of the piezocaloric
temperature change of K(H$_{1 - x}$D$_{x})_{2}$PO$_{4}$ crystals
at $x=0.0$ (solid lines) and $x=0.89$ (dashed lines) at
$T-T_c^0=-5$~K -- 1, $T=T_c^0$ -- 2,  $T-T_c^0=5$~K -- 3.}
\label{DTpk_KDP_DKDP}
\end{figure}

\begin{figure}[!h]
\begin{center}
 \includegraphics[scale=0.65]{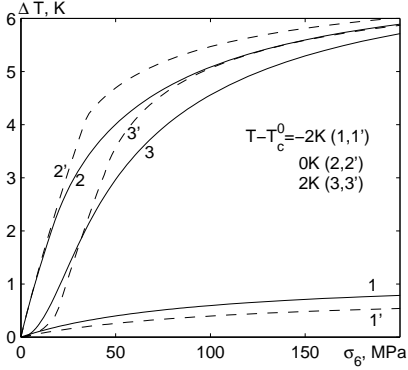}
\end{center}
\caption[]{The same of KH$_{2}$AsO$_{4}$ (solid lines) and KD$_{2}$AsO$_{4}$ (dashed lines) at
 $T-T_c^0=-2$~K -- 1, $T=T_c^0$ -- 2,  $T-T_c^0=2$~K -- 3.} \label{DTpk_KDA_DKDA}
\end{figure}

The calculated dependence of the piezocaloric temperature change $\Delta T$ of the
studied crystals on the adiabatically applied stress $\sigma_6$ is shown in figs.~\ref{DTpk_KDP_DKDP} and \ref{DTpk_KDA_DKDA}. %

At small $\sigma_6$ the piezocaloric temperature change increases
with the stress linearly at temperatures below $T_c^0$ (curves 1,
1') and quadratically above $T_c^0$ (curves 3, 3'). Here $T_c^0$
 denotes the transition temperature in absence of external fields.
 At $T=T_c^0$ and $\sigma_6<\sigma^*_6$ the temperature change is
equal to the shift of the transition temperature, which is roughly
proportional to the stress (see fig.~\ref{S_10MPa}). It is well
seen at $\sigma_6<40$~MPa in the KD$_{2}$AsO$_{4}$ crystal,
undergoing the most pronounced first order phase transition
(fig.~\ref{DTpk_KDA_DKDA}, curve 2'). In KH$_{2}$PO$_{4}$, where
the phase transition in zero fields is close
   to the second order, at $T=T_c^0$ and small stresses $\sigma_6<10$~MPa the temperature change follows the law $\Delta T \sim \sigma_6^{2/3}$.
   At large stresses the piezocaloric stress above $T_c^0$ is larger than below it, and the deviation from the linear or quadratic behavior
   is observed; at $\sigma_6 \gg 200$~MPa the saturation is reached. Unfortunately, no experimental data for $\Delta T$ is available.

\begin{figure}[!h]
\begin{center}
 \includegraphics[scale=0.7]{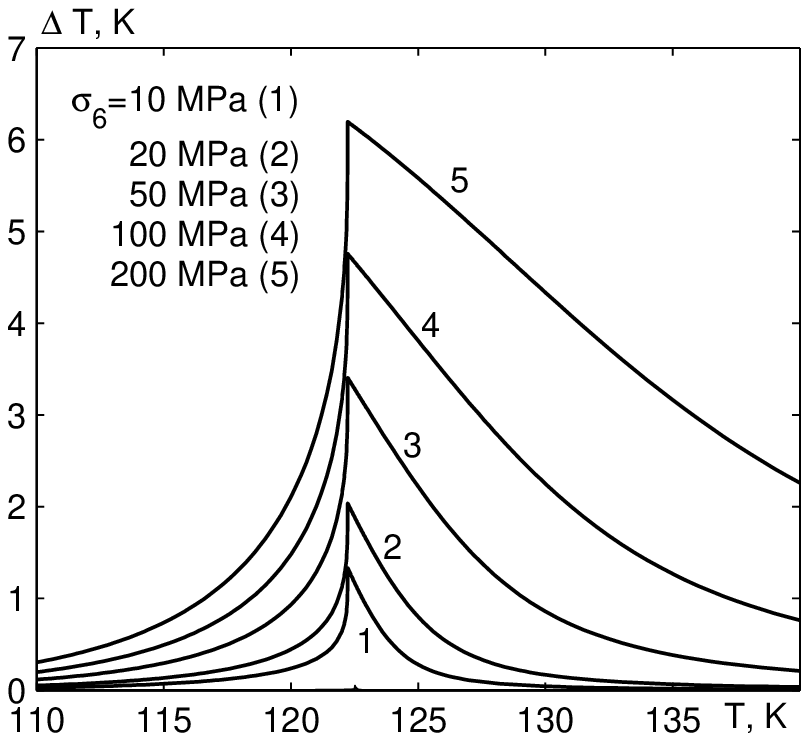}~~~\includegraphics[scale=0.7]{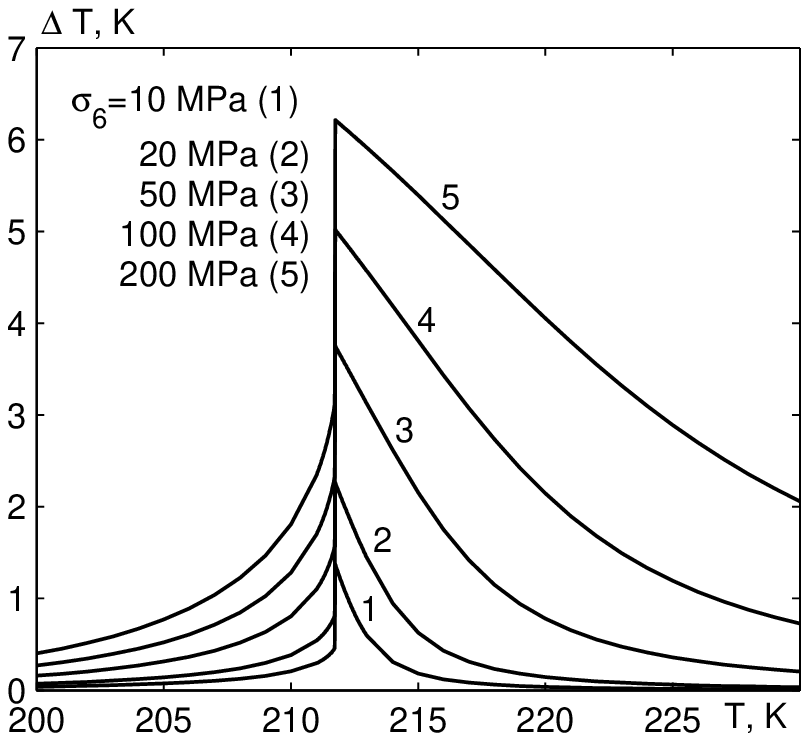}
\end{center}
\caption[]{The temperature dependence of the piezocaloric temperature change of K(H$_{1 - x}$D$_{x})_{2}$PO$_{4}$ at $x=0.0$ (left) and $x=0.89$
  (right) at different values of the stress $\sigma_6$.} \label{DTpk_T_x89}
\end{figure}

As follows from the temperature dependence of $\Delta T$ (fig.~\ref{DTpk_T_x89}), the
largest piezocaloric temperature change is expected just above $T_c$, where it can exceed 5~K.

\begin{figure}[!h]
\begin{center}
 \includegraphics[scale=0.7]{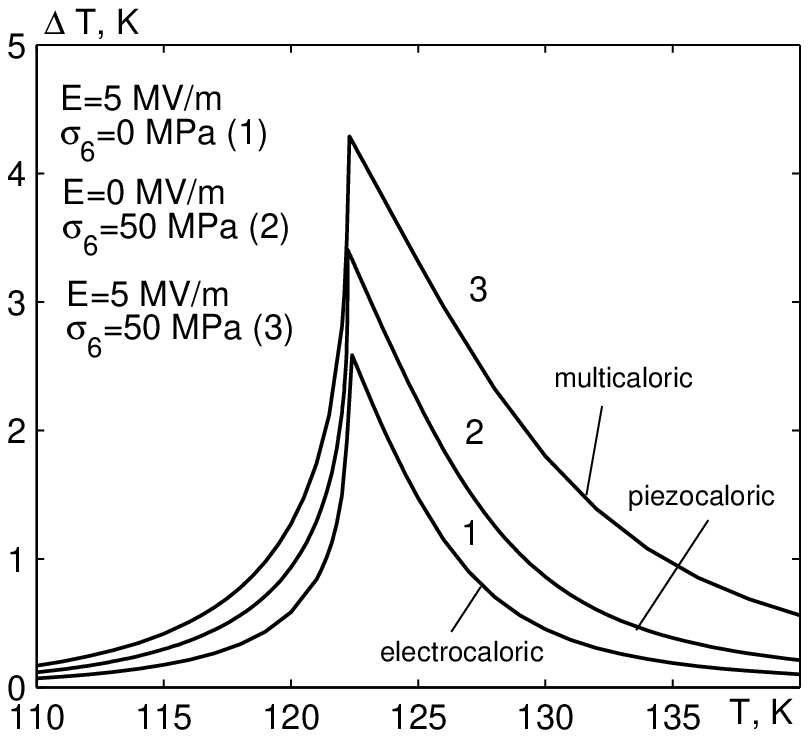}~~~\includegraphics[scale=0.7]{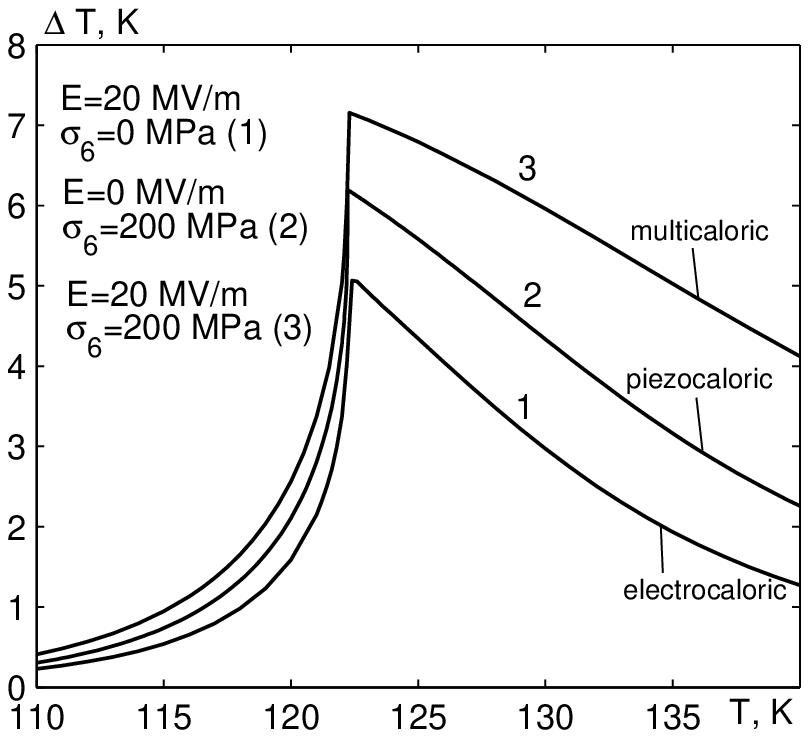}
\end{center}
\caption[]{The temperature dependence of the temperature change of the KH$_{2}$PO$_{4}$ crystal
under the electric field $E_3$ \cite{Vdovych2014} (curves 1), stress $\sigma_6$  (curves 2), the field and stress simultaneously (curves 3).}
\label{DT_p200_MPa_E200kVcm}
\end{figure}

Simultaneous application of the electric field $E_3$ and stress $\sigma_6$ yields the multicaloric effect (fig.~\ref{DT_p200_MPa_E200kVcm}).
It is stronger than either electrocaloric or piezocaloric effect, and
at temperatures far from $T_c^0$ the multicaloric temperature shift is even larger than the simple sum of the respective electro- and piezocaloric
shifts.

 \section{Conclusions}
Modification of the proton ordering model, suggested earlier in order to describe the electrocaloric effect in the KH$_{2}$PO$_{4}$ family crystals,
is used to explore the piezocaloric and multicaloric effects. They are found to be qualitatively and quantitatively similar to the
electrocaloric effect. The theory predicts that the temperature change can exceed 5~K at high stresses (obviously well above the sample strength)
and can be increased further by a simultaneous application of both electric field and shear stress $\sigma_6$. Experimental measurements of
$\Delta T$ are required.


\begin{thebibliography}{999}


\bibitem{Cui2012} J. Cui, Y. Wu, J. Muehlbauer, Y. Hwang, R. Radermacher,
S. Fackler, M. Wuttig,  I. Takeuchi, Appl. Phys. Lett.
\textbf{101}, 073904 (2012).

\bibitem{Pieczyska2006} E.A. Pieczyska, S.P. Gadaj, W.K. Nowacki, H. Tobushi, Experimental
Mechanics \textbf{46}, 531 (2006).

\bibitem{Bonnot2008} E. Bonnot, R. Romero, L. Manosa, E. Vives, A.
Planes, Phys. Rev. Lett. \textbf{100}, 125901 (2008).


\bibitem{Lisenkov2013} S. Lisenkov, B. K. Mani, C.-M. Chang, J. Almand, I.
Ponomareva, Phys. Rev. B. \textbf{87}, 224101 (2013).


\bibitem{Stasyuk2000}
Stasyuk I.V., Levitskii R.R., Zachek I.R., Moina A.P., Phys. Rev.
B \textbf{62}, 6198 (2000).

\bibitem{Stasyuk2001}
Stasyuk I.V., Levitskii R.R., Moina A.P., Lisnii B.M.,
Ferroelectrics \textbf{254}, 213 (2001).

\bibitem{lis2007}
 B.M. Lisnii, R.R. Levitskii, O.R. Baran, Phase Transitions \textbf{80}, 25 (2007).

\bibitem{JPS1701}   Levitsky R.R., Zachek I.R., Vdovych A.S., Moina A.P, J. Phys. Studies
\textbf{14}, 1701 (2010).



\bibitem{Vdovych2014prep} A.S. Vdovych, A.P. Moina, R.R. Levitskii, I.R. Zachek.// arXiv:1405.1327v1 [cond-mat.mtrl-sci] 6 May 2014.

\bibitem{Vdovych2014} A.S. Vdovych, A.P. Moina, R.R. Levitskii, I.R. Zachek, Condens. Matter Phys. \textbf{17}, 43703 (2014).







\end{thebibliography}
\end{document}